\documentclass[12pt]{article}

\def\vp{\varphi}

\def\half{\textstyle{\frac{1}{2}}}

\def\l{\lambda}

\def\ra{\rightarrow}
\def\tint{{\textstyle\int}}
\def\hg{{\hat g}}
\def\hp{{\hat\pi}}

\def\s{\hskip.08em}
\def\d{\partial}

\def\b{\begin{eqnarray*}}  
\def\e{\end{eqnarray*}}    
\def\bn{\begin{eqnarray}}  
\def\en{\end{eqnarray}}   

\def\<{\langle}
\def\>{\rangle}

\def\no{\nonumber}

\def\{{\lbrace}

\def\}{\rbrace}
\begin{document}

\title{A Simple Factor in \\Canonical Quantization yields \\Affine Quantization \\
Even for Quantum Gravity}
  \author{John R. Klauder\footnote{klauder@ufl.edu} \\
Department of Physics and Department of Mathematics  \\ 
University of Florida,   
Gainesville, FL 32611-8440}
\date{ }

\maketitle 
\begin{abstract}Canonical quantization (CQ) is built around $[Q,P]=i\hbar1\!\!1$,
while affine quantization (AQ) is built around $[Q,D]=i\hbar\,Q$, where $D\equiv(PQ+QP)/2$.
The basic CQ operators must fit $-\infty< P, Q <\infty$, while the basic AQ operators can fit 
$-\infty<P<\infty$ and $ 0<Q<\infty$, $-\infty <Q<0$, or even $-\infty<Q\neq0<\infty$.
AQ can also be the key to quantum gravity, as our simple outline demonstrates.
\end{abstract}
	\section{Canonical Quantization to \\Affine Quantization: CQ to AQ}
	The CQ equation $[Q,P]=QP-PQ=i\hbar 1\!\!1$ is {\bf multiplied} by $Q$ to give $\{Q(QP-PQ) +(QP-PQ)Q\}/2
= \{ Q^2P+QPQ-QPQ-PQ^2\}/2=[Q,(QP+PQ)/2]\equiv [Q,D]=i\hbar\,Q$. This principal equation {\bf requires} that 
$Q\neq0$, leading to the last expression which applies for $Q>0$, $Q<0$, or both. From a classical 
promotion viewpoint for CQ it is $-\infty <p\ra P, q\ra Q<\infty$, while for AQ it is
$-\infty <d\equiv pq\ra D\equiv (PQ+QP)/2<\infty$, while $-\infty< q\neq0\ra Q\neq0<\infty$. Note:
$P^\dag\neq P$, but $P^\dag Q=PQ$.

Not every set of classical variables lead to valid quantum operators, neither for CQ or for AQ.The 
correct choice of classical variables to promote to quantum operators for CQ are Cartesian coordinates, which creat a constant vanishing curvature. The classical coordinates that creat a constant negative curvature\footnote{A constant negative curvature is discussed in \cite{yy}.}   are the correct coordinates to promote to quantum operators for AQ.\footnote{
 Classical variables for spin variables that form a constant positive curvature are the final
 set of correct classical variables that promote to valid quantum operators. The constant (+, 0, -1)
 curvatures are all addressed in \cite{cnc}, Sec.~1.}
{\it Regarding CQ and AQ operators, the principal difference is that CQ operators must span the whole real line, while the AQ operator $Q$ strictly spans all positive coordinates, all negative coordinates, or both, provided $Q\neq 0$.}\footnote{Instead of $-\infty<Q\neq 0<\infty$, we can also have $-\infty<Q\neq b<\infty$, provided $b$ is finite.}

Different sets of classical problems require valid CQ or AQ quantizations. For example, a traditional harmonic oscillator, with $-\infty< p, q<\infty$, employs CQ, while a half-harmonic oscillator, with $-\infty< p<\infty$, but $0<q<\infty$, employs AQ \cite{lg}. Partial-harmonic oscillators, with $-\infty<p<\infty$, while $-b< q<\infty$ and $0<b<\infty$, also employ AQ \cite{hk}.

Examples from scalar field theories show that $\vp^{12}_3$ leads to ``free" results for CQ while
``non-free'' results for AQ \cite{r1}, and also ``free'' results for $\vp^4_4$ \cite{a,b,c} while AQ leads to a ``non-free'' result \cite{prd}. {\it It becomes clear that CQ is {\bf not} the only quantization procedure for various classical systems!} The author has even suggested that quantum field theory might benefit by adding AQ to the usual procedures \cite{67}.

 Regarding gravity, an affine result can offer a {\it strictly positive metric},  i.e., $ds(x)^2=
g_{ab}(x)\,dx^a\,dx^b>0$, provided $dx^c \equiv\hskip-0.93em/\;\,0$. This ability 
offers a strong contribution to the quantization of gravity, as sketched below.

\section{The Essence of Quantum Gravity}
The author has shown that an affine quantization is the natural procedure to quantize Einstein's               gravity, e.g., \cite{cnc, k3}. Here we offer an easy and straightforward 
derivation that features the highlights of a natural quantum gravity. Our procedure is based on the analysis of other quantum procedures that lead toward a conventual Hamiltonian as part of a standard Schr\"odinger's representation followed by a suitable version of Schr\"odinger's equation.

\subsection{Classical gravity}
The classical Hamiltonian, according to ADM \cite{adm}, uses the metric $g_{ab}(x)=g_{ba}(x)$ and its determinant $g(x)\equiv \det[g_{ab}(x)]$,\footnote{The additional symbol $\{g\}$ stands for $\{g_{ab}(\cdot)>0\}$.} the momentum 
$\pi^{cd}(x)=\pi^{dc}(x)$, or more useful, the momentric\footnote{The name momentric is taken from  {\it momen}tum and me{\it tric}.} $\pi^c_d(x)\equiv \pi^{ce}(x)\,g_{de}(x)$, all of which leads to
  \bn &&H(\pi^a_b, g_{cd})=\tint \{ g(x)^{-1/2}[\pi^a_b(x)\pi^b_a(x)-\half \pi^a_a(x)\pi^b_b(x)]\no\\
  &&\hskip8em +g(x)^{1/2}(x)\,R(x)\}\;d^3\!x \;, \en
  where $R(x)$ is the Ricci scalar \cite{7}. The Poisson brackets for the momentric and metric fields are given by
      \bn &&\{\pi^a_b(x),\pi^c_d(x')\}=
   \half\,\delta^3(x,x')\s[\delta^a_d\s \pi^c_b(x)-\delta^c_b\s \pi^a_d(x)\s]\;,    \no \\
       &&\hskip-.20em\{g_{ab}(x), \s \pi^c_d(x')\}= \half\,\delta^3(x,x')\s [\delta^c_a g_{bd}(x)+\delta^c_b g_{ad}(x)\s] \;,      \\
       &&\hskip-.30em\{g_{ab}(x),\s g_{cd}(x')\}=0 \;. \no  \en
       Observe that these Poisson brackets are valid even if we change $g_{ab}(x)$ to $-g_{ab}(x)$, a property that allows us to retain only $g_{ab}(x)>0$. This is not possible with the Poisson brackets for canonical variables.
      
  \subsection{Quantum variables}
  Passing to operator commutations by a  
 promotion of the set of Poisson brackets to operator commutations leads, as established in \cite{k3}, to 
 \bn   &&[\hp^a_b(x),\s \hp^c_d(x')]=i\s\half\,\hbar\,\delta^3(x,x')\s[\delta^a_d\s \hp^c_b(x)-\delta^c_b\s \hp^a_d(x)\s]\;,    \no \\
       &&\hskip-.10em[\hg_{ab}(x), \s \hp^c_d(x')]= i\s\half\,\hbar\,\delta^3(x,x')\s [\delta^c_a \hg_{bd}(x)+\delta^c_b \hg_{ad}(x)\s] \;, \\
       &&\hskip-.20em[\hg_{ab}(x),\s \hg_{cd}(x')] =0 \;. \no  \en
As with the Poisson brackets, these commutators are valid if we change  $\hg_{ab}(x)$ to $-\hg_{ab}(x)$. For the momentric and metric operators,
we again find that we can retain only $\hg_{ab}(x)>0$. Moreover,  with suitable regularization, it follows that $\hp^a_b(x)\,\hg(x)^{-1/2}=0$, which is proved in  \cite{k3}.

\subsection{Schr\"odinger's representation and equation}
Schr\"odinger's representation chooses $\hg_{ab}(x)= g_{ab}(x)$ and
 \bn \hp^c_d(x) =  -i\half \hbar
[g_{de}(x)(\d/\d g_{ce}(x))+(\d/\d g_{ce}(x)) g_{de}(x)] \;, \en
which is well evaluated using a suitable regularization \cite{k3}.
Finally, we are led to Schr\"odinger's
equation 
   \bn &&\hskip-2em i\hbar \, \d \;\Psi(\{g\},t)/\d t= \{\!\!\{\tint\{ [\hp^a_b(x) \,g(x)^{-1/2}\,\hp^b_a(x)-\half \hp^a_a(x)\,
   g(x)^{-1/2} \,\hp^b_b(x)] \no \\
   &&\hskip10em  +g(x)^{1/2}\, R(x) \}\;d^3\!x \}\!\!\}\;\Psi(\{g\},t)\;. \en

   \subsection{Potential wave functions}
   The expression $\hp^a_b(x)\,g(x)^{-1/2}=0$ points to a form of wave function given by
       $ \Psi(\{g\})=W(\{g\}) \Pi_y \,g(y)^{-1/2}$. The wave functions for the
       Hamiltonian are continuous in 
       the metrics that make up $\{g\}$, which is natural for a Hamiltonian that contains 
       spatial derivatives of the metric fields and enforces continuity within the Ricci 
       scalar, $R(x)$. 
       
       The normalization of such wave functions may be given by
       \bn \tint |\Psi(\{g\})|^2 {\cal{D}}\{g\}= \tint |W(\{g\})|^2 \Pi_y\,g(y)^{-1}\:{\cal{D}}
       \{g\} =1 \;, \en
       while a matrix expression for an operator $A$ may be given by 
       \bn &&\tint\Psi'(\{g'\})^* \,A(\{g'\},\{g\})\,\Psi(\{g\})\;{\cal{D}}\{g'\}\,{\cal{D}}\{g\}\no\\
       &&\hskip2em=\tint\Pi_{y'} g'(y')^{-1/2} A(\{g'\},\{g\}) \/\Pi_yg(y)^{-1/2}\;{\cal{D}}\{g'\}\,
       {\cal{D}}\{g\} \;. \en
       It may be noticed that the factors $\Pi_y\,g(y)^{-1/2}$ act somewhat like `coherent states',           
        e.g., \cite{bs}, in these equations.

       \section{Conclusion}
       
       The forgoing study is focussed on the gravity Hamiltonian, which is the most difficult aspect of quantum gravity. Additional topics, such as fulfilling constraints, are examined in \cite{k3}, along with a somewhat different viewpoint in \cite{cnc}. An earlier paper \cite{ee} is especially focussed on establishing a path through the multiple constraints that quantum gravity faces.
       
       Contributions by other researchers using affine procedures to quantize gravity are  welcome to add to the present story.


\begin{thebibliography}{99}

\bibitem{yy} ''constant negative curvature``; Scholarpedia: Hyperbolic dynamics

\bibitem{cnc} J. Klauder, ``Using Coherent States to Make Physically Correct Classical-to-Quantum Procedures That Help Resolve Nonrenomalizable Fields Including Einstein’s Gravity", {\it Journal of High Energy Physics, Gravitation and Cosmology} {\bf 7}, (1019-1026 (2021); 
DOI:10.4236/jhepgc.2021.73060.

\bibitem{lg} L. Gouba, ``Affine Quantization on the Half Line'',   {\it Journal of High Energy Physics, Gravitation and Cosmology} {\bf 7}, 352-365 (2021); DOI:10.4236/jhepgc.2021.71019. 

\bibitem{hk} C. Handy and J. Klauder, ``Proof that Half-Harmonic Operators become Full-Harmonic Oscillators after the Wall Slides Away'',   arXiv:2108.00289.



\bibitem{r1} R. Fantoni, ``Monte Carlo Evaluation of the Continuum Limit of $\vp^{12}_3$'',
J. Stat. Mech. (2021) 083102; arXiv:2011.09862.

    
\bibitem{a} B. Freedman, P. Smolensky, D. Weingarten, ``Monte Carlo Evaluation of the Continuum
Limit of $\varphi^4_4$ and $\varphi^4_3$'', Phys. Lett. B {\bf 113}, 481 (1982). 

    \bibitem{b} M. Aizenman, ``Proof of the Triviality.  of                       
$\varphi^4_d$ Field  Theory and Some Mean-Field Features of Ising
Models for $d>4$", Phys. Rev. Lett.{\bf 47}, 1-4, E-886
(1981).

\bibitem{c} J. Fr\"ohlich, ``On the Triviality of $\l\varphi^4_d$             
Theories and the Approach to the Critical Point in $d\ge 4$
Dimensions'', { Nuclear Physics B} {\bf 200}, 281-296 (1982).


\bibitem{prd} R. Fantoni and J. Klauder, ``Affine Quantization of $\vp^4_4$ Succeeds While Canonical Quantization Fails'', Phys. Rev. D {\bf 103}, 076013 (2021).
 
 \bibitem{67} J. Klauder, ``Evidence for Expanding Quantum Field Theory``,{\it Journal of High Energy Physics, Gravitation and Cosmology} {\bf 7}, 1157-1160 (2021); DOI:10.4236/jhepgc.2021.73067.


\bibitem{k3} J. Klauder, ``Using Affine Quantization to Analyze Nonrenotmalizable Scalar Fields and the Quantization of Einstein’s Gravity'', {\it Journal of High Energy Physics, 9Gravitation and Cosmology} {\bf 6}, 802-816 (2020); DOI:10.4236/jhepgc.2020.64053.


\bibitem{adm} R.  Arnowitt, S. Deser, and C. Misner,  ``The Dynamics of General Relativity'', {\it Gravitation: An Introduction to 
Current Research}, Ed. L. Witten, (Wiley \& Sons, New York, 1962), p. 227; arXiv:gr-qc/0405109.

\bibitem{7} ``Ricci scalar'': Wikipedia; https;//wikipedia Ricci scalar

\bibitem{bs} J. Klauder and B.-S. Skagerstam,  (1985) ``Coherent States: Applications in Physics and Mathematical Physics'', World Scientific, Singapore; DOI:/org/10.1142/0096
 
 \bibitem{ee} J. Klauder, ``Fundamentals of Quantum Gravity'', arXiv:gr-qc/0612168v1.


\end{thebibliography}
\end{document}